\documentclass[twocolumn]{aastex631}

\usepackage{mathrsfs}
\usepackage{graphicx}
\usepackage{rotating}
\usepackage{amsmath}
\usepackage{bm}
\usepackage{ulem}
\usepackage{CJK}

\DeclareMathOperator{\sech}{sech}

\shortauthors{Hozumi, Nitadori, \& Iwasawa}

\begin{document}
\begin{CJK*}{UTF8}{ipxm}
\hyphenation{Poisson Athanassoula Iwasawa Sellwood Rubin Hernquist
Ostriker Plummer}

\title{SCF-FDPS: A Fast $N$-body Code for Simulating Disk-halo Systems}

\correspondingauthor{Shunsuke Hozumi}
\email{hozumi@edu.shiga-u.ac.jp}

\author[0000-0003-2357-881X]{Shunsuke Hozumi (穂積俊輔)}
\affiliation{Faculty of Education, Shiga University,
2-5-1 Hiratsu, Otsu, Shiga 520-0862, Japan}

\author{Keigo Nitadori (似鳥啓吾)}
\affiliation{RIKEN Center for Computational Science,
7-1-26 Minatojima-minami-machi, Chuo-ku, Kobe, Hyogo 650-0047, Japan}

\author{Masaki Iwasawa (岩澤全規)}
\affiliation{Department of Information Engineering, National Institute of
Technology, Matsue College, 14-4 Nishi-Ikuma Cho, Matsue, Shimane
690-8518, Japan}

\begin{abstract}
A fast $N$-body code has been developed for simulating a stellar disk
embedded in a live dark matter halo.  In generating its Poisson solver,
a self-consistent field (SCF) code which inherently possesses perfect
scalability is incorporated into a tree code which is parallelized
using a library termed Framework for Developing Particle Simulators
(FDPS).  Thus, the code developed here is called SCF-FDPS.  This code
has realized the speedup of a conventional tree code by applying
an SCF method not only to the calculation of the self-gravity of the
halo but also to that of the gravitational interactions between the
disk and halo particles.  Consequently, in the SCF-FDPS code, a tree
algorithm is applied only to calculate the self-gravity of the disk.
On a many-core parallel computer, the SCF-FDPS code has performed
at least three times, in some case nearly an order of magnitude,
faster than an extremely-tuned tree code on it, if the numbers
of disk and halo particles are, respectively, fixed for both codes.
In addition, the SCF-FDPS code shows that the cpu cost scales almost
linearly with the total number of particles and almost inversely
with the number of cores.  We find that the time evolution of
a disk-halo system simulated with the SCF-FDPS code is, in large
measure, similar to that obtained using the tree code.  We suggest
how the present code can be extended to cope with a wide variety
of disk-galaxy simulations.

\end{abstract}


\section{Introduction}\label{sec:introduction}
The number of particles in $N$-body simulations of astronomical
objects like galaxies has been increasing, in step with the
progress in parallel computing technology.  This remarkable
development has brought a great benefit to disk-galaxy simulations,
because galactic disks are rotation-supported, cold systems, so
that a sufficiently large number of particles are needed for the
disk to sidestep the heating originating from Poisson noise.
In fact, \citet{fujii11} have demonstrated that a spiral
feature emerging in a disk surrounded by an unresponsive
halo is fading away gradually over time for a million-particle
simulation, while it persists until late times for
a three-million-particle simulation.  On the other hand,
\citet{lia02} has revealed that for a given disk-halo model,
the disk is stabilized against bar formation when the halo
is rigid, while a large-amplitude bar is excited through a
wave-particle resonance between a bar mode in the disk and
halo particles when the halo is live.  This fact coerces us
to deal with a halo as self-gravitating.  In making a halo
live for a disk-halo system, the mass of a halo particle has
to be made equal to that of a disk particle to avoid the shot
noise generated by halo particles when they pass through the
disk.  Unfortunately, a halo mass is estimated to be at least
around an order of magnitude larger than a disk mass, because
a halo is considered to extend far beyond the optical edge of
the disk on the basis of the observed rotation curves of disk
galaxies that are, in general, flat out to large radii
\citep[e.g.,][]{sofue01}.  Consequently, the number of
halo particles becomes larger than that of disk particles
by an order of magnitude or many more.  It thus follows
that disk-galaxy simulations inevitably demand a large
number of particles.

As the number of particles in $N$-body simulation increases, the
number of force calculation increases accordingly.  Because a given
particle receives the gravitational force from all other particles
in an $N$-particle system, the number of the total force calculation
reaches $\mathcal{O}(N^2)$ at every time step in the simplest
way.  This explosive nature in force calculation is alleviated
down to $\mathcal{O}(N\log N)$ by the introduction of a tree
algorithm developed by \citet{bh86}.  Indeed, recent large $N$-body
simulations of disk galaxies are based on a tree code.  For example,
\citet{dbs09} adopted a parallelized tree code to investigate the
bar instability in galactic disks using $1.8\times 10^7$ particles
for a disk and $10^8$ particles for a halo, while \citet{dvh13}
used a tree-based gravity solver to examine the origin of spiral 
structure in disk galaxies with $10^8$ particles for a disk
immersed in a rigid halo.  Furthermore, \citet{fujii18} have
employed a tree-based code called BONSAI \citep{bedorf12}
optimized for Graphics Processing Units to scrutinize the
dynamics of disk galaxies which consist of live disk, bulge,
and halo components with the total number of particles being
increased up to $5\times 10^8$.  In their subsequent work,
\citet{fujii19} have boosted the total number of particles
up to $8\times 10^9$ to construct the Milky Way Galaxy model
that reproduces the observed properties.

As mentioned above, a tree algorithm is commonly used to study
disk galaxies with a huge number of particles.  In such a situation,
a faster tree code is understandably desirable from various aspects
of numerical studies.  As computer architecture is shifted to parallelized
one, a tree code has been adjusted to a parallel computer.  Above all,
a numerical library termed Framework for Developing Particle Simulators
(FDPS) \citep{iwasawa16, namekata18} has tuned a tree code to the utmost
limit of a massively memory-distributed parallel computer.  Therefore,
no further speedup of existing tree codes is expected on their own.

We then try to incorporate a self-consistent field (SCF) code
into a tree code.  Of course, the FDPS library is implemented
in the tree part of the resulting hybrid code for the efficient
parallelization.  In an SCF approach, Poisson's equation is solved
by expanding the density and potential of the system being studied
in a set of basis functions.  In particular, owing to the expansion
of the full spatial dependence, the cpu cost becomes $\mathcal{O}(N)$.
Moreover, because the perfect scalability is inherent in the SCF
approach, it is suitable for parallel computing.  By taking advantage
of these characteristics, we will be able to accelerate $N$-body
simulations of disk galaxies using a hybrid code named SCF-FDPS
in which an SCF code is incorporated into an FDPS-implemented
tree code \citep{shunsuke_hozumi_2023_7633122}.

In this paper, we describe how an SCF code is incorporated into
a tree code, and show how well the resulting SCF-FDPS code works.
In Section~\ref{sec:models}, we present the details of the SCF-FDPS
code, including how an SCF approach is applied to a disk-halo system.
In Section~\ref{sec:tests}, along with the determination of the
parameters inherent in the code, the performance of the code is
shown.  In Section~\ref{sec:discussion}, we discuss the extension
of the present code to cope with a wide variety of disk-galaxy
simulations.  Conclusions are given in Section~\ref{sec:conclusions}.

\section{Details of the SCF-FDPS Code}\label{sec:models}
We develop a fast $N$-body code which is based on both SCF and
tree approaches.  First, we explain the SCF method briefly, and
then, describe the details of the SCF-FDPS code.

\subsection{SCF Method}
An SCF method requires a biorthonormal basis set which satisfies
Poisson's equation written by
\begin{equation}
    \nabla^2\Phi_{nlm}(\bm r)=4\pi G\rho_{nlm}(\bm r),
\end{equation}
where $\rho_{nlm}(\bm r)$ and $\Phi_{nlm}(\bm r)$ are,
respectively, the density and potential basis functions at
the position vector of a particle, $\bm r$, with $n$ being
the `quantum' number in the radial direction and with $l$
and $m$ being corresponding quantities in the angular directions.
Here, the biorthonormality is represented by
\begin{equation}
    \int \rho_{nlm}(\bm r)\,[\Phi_{n'l'm'}(\bm r)]^*
    d\bm r=\delta_{nn'}\delta_{ll'}\delta_{mm'},
\label{orthonormality}
\end{equation}
where $\delta_{kk'}$ is the Kronecker delta defined by
$\delta_{kk'}=0$ for $k\ne k'$ and $\delta_{kk'}=1$ for
$k=k'$.

With the help of such a biorthonormal basis set
as is noted above, the density and potential of the system
are expanded, respectively, by the corresponding basis
functions as
\begin{equation}
    \rho(\bm r)=\sum_{n,l,m} \,A_{nlm} \, \rho_{nlm}(\bm r)
    \label{density}
\end{equation}
and
\begin{equation}
    \Phi(\bm r)=\sum_{n,l,m} A_{nlm} \, \Phi_{nlm}(\bm r),
    \label{potential}
\end{equation}
where $A_{nlm}$ are the expansion coefficients at \mbox{time $t$}.
When the potential basis functions are operated on the density
field that is expanded as Equation~(\ref{density}), $A_{nlm}$
are given, via the biorthonormality relation of
Equation~(\ref{orthonormality}), by
\begin{equation}
    A_{nlm}=\int \rho(\bm r)\,[\Phi_{nlm}(\bm r)]^*\,d\bm r.
    \label{int_coef}
\end{equation}
If a system consists of a collection of $N$ discrete mass-points,
the density is represented by
\begin{equation}
\rho(\bm r)=\sum_{k=1}^N m_k\,\delta(\bm r-\bm r_k),
\label{discrete_density}
\end{equation}
so that by substituting Equation~(\ref{discrete_density}) into
Equation~(\ref{int_coef}), $A_{nlm}$ result in
\begin{eqnarray}
A_{nlm} & = & \int \sum_{k=1}^Nm_k\,
\delta(\bm r-\bm r_k)\,[\Phi_{nlm}(\bm r)]^*\,d\bm r\nonumber\\
& = & \sum_{k=1}^N\,m_k[\Phi_{nlm}(\bm r_{k})]^*,
\label{sum_coef}
\end{eqnarray}
where $m_k$ and $\bm r_{k}$ are the mass and position
vector of the $k$th particle in the system, respectively,
and $\delta(\bm r)$ is Dirac's delta function.  After obtaining
$A_{nlm}$, we can derive the acceleration, $\bm a(\bm r)$, by
differentiating Equation~(\ref{potential})
with respect to $\bm r$, finding
\begin{equation}
    \bm a(\bm r) = -\sum_{nlm} A_{nlm}\,\nabla\Phi_{nlm}(\bm r),
    \label{accel}
\end{equation}
where $\nabla\Phi_{nlm}(\bm r)$ can be analytically calculated
beforehand, once the basis set is specified.

As found from Equation~(\ref{sum_coef}), this form of summation can
be conveniently parallelized, so that an SCF code realizes the perfect
scalability \citep{hsb95}, which leads to ideal load balancing on a
massively parallel computer.  In addition, the cpu time is proportional to
$N\times$\mbox{$(n_\text{\rm max}+1)$}$\times$\mbox{$(l_\text{\rm max}+1)^2$},
where $n_\text{\rm max}$ and $l_\text{\rm max}$ are the maximum numbers of
expansion terms in the radial and angular directions, respectively.  Therefore,
an SCF code is fast and suitable for modern parallel computers.  Accordingly,
a fast $N$-body code is feasible by incorporating an SCF code into a tree code.

\subsection{The SCF-FDPS Code}
For a disk-halo system, the acceleration of the $k$th disk particle,
$\bm a_\text{d}(\bm r_{\text{d}, k})$,
at the position vector,
$\bm r_{\text{d}, k}$, and the acceleration of the $k$th halo particle,
$\bm a_\text{h}(\bm r_{\text{h},k})$,
at the position vector,
$\bm r_{\text{h}, k}$, are, respectively, represented by
\begin{equation}
\bm a_\text{d}(\bm r_{\text{d},k}) = 
\bm a_{\text{d} \to \text{d}}(\bm r_{\text{d},k}) + 
\bm a_{\text{h} \to \text{d}}(\bm r_{\text{d},k})
\label{accel_disk}
\end{equation}
and
\begin{equation}
\bm a_\text{h}(\bm r_{\text{h},k}) = 
\bm a_{\text{h} \to \text{h}}(\bm r_{\text{h},k}) + 
\bm a_{\text{d} \to \text{h}}(\bm r_{\text{h},k}),
\label{accel_halo}
\end{equation}
where $\bm a_{\text{d} \to \text{d}}(\bm r_{\text{d},k})$
and $\bm a_{\text{h} \to \text{d}}(\bm r_{\text{d},k})$
denote the acceleration due to the gravitational force from
other disk particles to the $k$th disk particle and that from
halo particles to the $k$th disk particle, respectively, while
$\bm a_{\text{h} \to \text{h}}(\bm r_{\text{h},k})$ and
$\bm a_{\text{d} \to \text{h}}(\bm r_{\text{h},k})$ stand
for the acceleration due to the gravitational force from
other halo particles to the $k$th halo particle and that from 
disk particles to the $k$th halo particle, respectively.

\citet{vs98} have already developed a code named {\sc scftree} in
which an SCF code is incorporated into a tree code.  In their code,
$\bm a_{\text{h} \to \text{h}}(\bm r_{\text{h},k})$
and
$\bm a_{\text{h} \to \text{d}}(\bm r_{\text{d},k})$
are calculated with an SCF method, while
$\bm a_{\text{d} \to \text{d}}(\bm r_{\text{d},k})$
and 
$\bm a_{\text{d} \to \text{h}}(\bm r_{\text{h},k})$
are manipulated with a tree method.  However, as explained in
Section~\ref{sec:introduction}, the number of halo particles
is at least about an order of magnitude larger than that of
disk particles, so that the calculation of
$\bm a_{\text{d} \to \text{h}}(\bm r_{\text{h},k})$
is extremely time-consuming, if a tree algorithm is used.  Of
course, local small-scale irregularities often generated in a
rotation-supported disk can be well-described with a tree code,
which makes it reasonable to apply a tree method to the calculation
of $\bm a_{\text{d} \to \text{d}}(\bm r_{\text{d},k})$.  In contrast,
in a halo which is supported by velocity dispersion, global features
survive but small-scale ones are smoothed out to disappear, so that
we can handle a halo using an SCF approach without so many expansion
terms.  In fact, there are suitable basis sets for spherical systems
whose density and potential are reproduced with a small number of
expansion terms.  Then, we apply an SCF method to evaluate
$\bm a_{\text{h} \to \text{h}}(\bm r_{\text{h},k})$.
Furthermore, even though small-scale features exist
in the disk, they do no serious harm to the overall
structure of the halo, as we will show in Section~\ref{sec:tests}.
Therefore, we can apply an SCF method to the calculation
of $\bm a_{\text{d} \to \text{h}}(\bm r_{\text{h},k})$
as well.  After all, only
$\bm a_{\text{d} \to \text{d}}(\bm r_{\text{d},k})$
is calculated with a tree method.  For this part in
the code, we implement a C++ version of the FDPS library
\citep{iwasawa16} which is publicly available, because
it helps users parallelize a tree part easily with
no efforts in tuning the code for parallelization.
We then name the code developed here the SCF-FDPS
code \citep{shunsuke_hozumi_2023_7633122}.  This
code will enable us to simulate disk-halo systems
much faster than ever for the fixed number of particles.

The actual procedure for calculating the accelerations of
$\bm a_{\text{h} \to \text{d}}(\bm r_{\text{d},k})$,
$\bm a_{\text{h} \to \text{h}}(\bm r_{\text{h},k})$, and
$\bm a_{\text{d} \to \text{h}}(\bm r_{\text{h},k})$ are
as follows.  First, Equation~(\ref{accel}) shows that
$\bm a_{\text{h} \to \text{d}}(\bm r_{\text{d},k})$
is provided by
\begin{equation}
\bm a_{\text{h} \to \text{d}}(\bm r_{\text{d},k})
=-\sum_{n,l,m} {A_{\text{h},nlm}\,
\nabla\Phi_{nlm}(\bm r_{{\text d},k}}),
\label{accel_hd}
\end{equation}
where $A_{\text{h},{nlm}}$ are those expansion coefficients
obtained from halo particles which are given by
\begin{equation}
A_{\text{h},nlm} = \sum_{k=1}^{N_\text{halo}}m_{\text{h},k}[\Phi_{nlm} (\bm r_{\text{h},k})]^*.
\label{eq:Ah_nml}
\end{equation}
In Equation~(\ref{eq:Ah_nml}), $N_\text{halo}$ is the number of halo
particles, and $m_{\text{h},k}$ is the mass of the $k$th halo particle.

Next, as is shown by Equation~(\ref{accel_halo}),
$\bm a_{\text{h} \to \text{h}}(\bm r_{\text{h},k})$
and $\bm a_{\text{d} \to \text{h}}(\bm r_{\text{h},k})$ are
added up to generate $\bm a_\text{h}(\bm r_{\text{h},k})$,
and again Equation~(\ref{accel}) indicates that
$\bm a_\text{h}(\bm r_{\text{h},k})$ is calculated as
\begin{equation}
\bm a_\text{h}(\bm r_{\text{h},k})
=-\sum_{n,l,m} {A_{\text{h+d},nlm}\,
\nabla\Phi_{nlm}(\bm r_{\text{h},k}}),
\label{accel_h}
\end{equation}
where $A_{\text{h+d},nlm}$ are those expansion coefficients
evaluated from disk and halo particles which are written by
\begin{equation}
A_{\text{h+d},nlm} = A_{\text{h},nlm} + A_{\text{d},nlm}.
\label{dhcoef}
\end{equation}
Here, $A_{\text{d},nlm}$ are the expansion coefficients
that are calculated from disk particles as
\begin{equation}
A_{\text{d},nlm} = \sum_{k=1}^{N_\text{disk}}m_{\text{d},k}[\Phi_{nlm}(\bm r_{\text{d},k})]^*,
\end{equation}
where $N_\text{disk}$ is the number of disk particles, and $m_{\text{d},k}$
is the mass of the $k$th disk particle.

In summary, the hybrid code is based on the following
Hamiltonian of the system written by
\renewcommand{\Re}{\operatorname{Re}}
\renewcommand{\Im}{\operatorname{Im}}
\begin{align}
H =& \sum_{k=1}^{N_\text{disk}} \frac{| \bm p_{\text{d},k} |^2} {2 m_{\text{d},k}} 
  + \sum_{k=1}^{N_\text{halo}} \frac{| \bm p_{\text{h},k} |^2} {2 m_{\text{h},k}}
  \nonumber \\
  &- \sum_{k=1}^{N_\text{disk}}  \sum_{j=k+1}^{N_\text{disk} }
    \frac{G {m_{\text{d},k}}{m_{\text{d},j}}}{\sqrt{| \bm r_{\text{d},k} - \bm r_{\text{d},j} |^2 + \varepsilon^2}}
  \nonumber \\
  &+ \frac12 \sum_{n,l,m} \sum_{k=1}^{N_\text{halo}} \sum_{j=1}^{N_\text{halo}} \nonumber \\
  &\qquad m_{\text{h},k} m_{\text{h},j}
     \Phi_{nlm}(\bm r_{\text{h},k}) \, [\Phi_{nlm}(\bm r_{\text{h},j})]^*
  \nonumber \\
  &+ \sum_{n,l,m} \sum_{k=1}^{N_\text{disk}} \sum_{j=1}^{N_\text{halo} } \nonumber\\ 
  & \qquad m_{\text{d},k} m_{\text{h},j} 
     \Re \left( 
     \Phi_{nlm}(\bm r_{\text{d},k}) \, [\Phi_{nlm}(\bm r_{\text{h},j})]^* 
     \right),
\end{align}
where
$\bm p_{\text{d},k} = m_{\text{d},k} \dot {\bm r}_{\text{d},k}$
and $\bm p_{\text{h},k} = m_{\text{h},k} \dot {\bm r}_{\text{h},k}$
are the momentum of the $k$th disk particle and that of the $k$th
halo particle, respectively.  The first two terms are kinetic ones.
The third term is the self-gravity of the disk that is calculated
with a tree method based on the softened gravity of the Plummer type
using a softening length, $\varepsilon$.  Notice that this expression
is used for convenience.  That is, it is incorrect in a strict sense,
because we cannot exactly construct the Hamiltonian owing to the way
of calculating the gravitational force in the tree algorithm.  The
fourth term is the self-gravity of the halo expressed by the expansions
due to the basis functions introduced into the SCF method.  The last
term represents the disk-halo interactions that are also expanded with
the basis functions.

We have postulated above that each particle in a disk-halo system has
a different mass.  In fact, the SCF-FDPS code supports individually
different masses for constituent particles in such a system.  However,
the mass of a halo particle should be made identical to that of a disk
particle so as to prevent the shot noise caused by the halo particles
that pass through the disk.  Consequently, in a practical sense, it is
appropriate to assign an identical mass to each particle in a disk-halo
system.

Now that the left-hand sides of Equations~(\ref{accel_disk}) and
(\ref{accel_halo}) are obtained as explained above, we can simulate
a disk-halo system with the code developed here.  As a cautionary
remark, we need a relatively large number of the angular expansion
terms to capture the gravitational contribution from disk particles
to halo particles properly, because the disk geometry deviates from
a spheroidal shape to a considerable degree.

\subsection{Parallelization}\label{subsec:parallel}
All simulations of the disk-halo system are run on a machine with
an AMD Ryzen Threadripper 3990X 64-core processor.  Although all 64
cores of this processor share the main memory, we apply not the thread
parallelization but the MPI parallelization to the SCF-FDPS code, and
execute simulations on up to 64 processes. 

The MPI parallelization of the SCF part in the SCF-FDPS code is
straightforward: once the particles are equally distributed to each
process, only one API call, {\tt MPI\_Allreduce()}, is needed for the
summation of those expansion coefficients which are calculated on each
process.  Regarding the SCF part, we do not need to move particles
across MPI processes.  On the other hand, the parallelization of the
tree part is more formidable than that of the SCF part, because we
have to take into consideration the spatial decomposition and exchange
of both particles and tree information between domains.  Fortunately,
the FDPS library copes with this complexity so as to be hidden from
the programmers.

\subsection{Hardware-specific Tuning}
The processor mentioned in Subsection~\ref{subsec:parallel} supports
up to 256-bit width SIMD instructions known as AVX.  This corresponds
to four words of double-precision numbers, or eight words of
single-precision numbers as the word length that can be processed
at once.  We conservatively adopt double-precision arithmetic
in the SCF-FDPS code to establish a reliable calculation method.
A further speedup by using the single-precision is the subject
of future work.  Thus, a speedup to a fourfold increase is expected
if the SIMD instructions are available.

In general, compiler's vectorization is applied to the innermost
loops.  However, this is not always the optimal way to exploit SIMD
instructions.  In the SCF-FDPS code, the compute kernel of the SCF
part consists of the outermost loop for the particle index $k$ and
several inner loops for the indices $n$, $l$, and $m$ that accompany
the basis functions.  Some of the inner loops can hardly be vectorized
because of their recurrence properties.  Thus, the maximal SIMD
instruction rate is achieved when the vectorization is applied to
the particle index $k$.  To this end, we write the compute kernel
of the SCF part in the SCF-FDPS code by the intrinsic functions
of AVX to manually vectorize the outermost loop.  In this way,
the positions and masses of four particles are fetched at once,
and the values of the basis functions are computed in parallel.

For the tree part, the compute kernel takes a double-loop form
composed of an outer loop for the sink particles that feel the
gravitational force and an inner loop for the source particles
that attract others.  Of the two  loops, the SIMD conversion is
applied to the outer loop through the intrinsic functions.  The
benefit of the outer-loop parallelization is the reduction of
memory access, because fetching the coordinates and mass of one
source particle to accumulate the gravitational forces for four
sink particles is more efficient than fetching four source particles
to accumulate  the gravitational forces to one sink particle.

\subsection{Portability}
As we have mentioned, the compute kernels of the SCF part  and
the tree part in the SCF-FDPS code are written using the intrinsic
functions of AVX.  However, that code can be compiled not only by
the Intel compiler but also by GCC and LLVM Clang.  At the same time,
it can run on other x86 processors which support AVX/AVX2.  Except in
the SIMD intrinsics, the SCF-FDPS code is written in standard C++17
and MPI, so that it runs on the arbitrary number of processors as
well as on the 64-core processors used here, regardless of whether
processors are configured within a node or shared over multiple nodes.
In fact, we have confirmed that the SCF-FDPS code can run properly
using 512 cores on a Cray XC50 system.

\section{Tests of the SCF-FDPS Code}\label{sec:tests}
\subsection{Disk-halo Model}\label{subsec:model}
We use a disk-halo model to examine the performance of the
SCF-FDPS code.  The disk model is an exponential disk
which is locally isothermal in the vertical direction.  The
volume density distribution, $\rho_\text{d}$, is given by
\begin{equation}
\rho_\text{d}(R,\,z)=\frac{M_\text{d}}{4\pi h^2z_0}
\exp(-R/h)\sech^2(z/z_0),
\label{eq:disk}
\end{equation}
where $R$ is the cylindrical radius, $z$ is the vertical
coordinate with respect to the mid-plane of the disk, $M_\text{d}$
is the disk mass, $h$ is the radial scale length, and $z_0$ is
the vertical scale length being set to be $0.2\,h$.  The disk
is truncated explicitly at $R=15\,h$ in the radial direction.
On the other hand, the halo model is described by an NFW profile
\citep{nfw96,nfw97}, whose density distribution,
$\rho_\text{\rm h}$, is written by
\begin{equation}
\rho_\text{h}(r)=\frac{\rho_0}{(r/r_\text{s})(1+r/r_\text{s})^2},
\label{eq:NFWhalo}
\end{equation}
where $r$ is the spherical radius, $r_\text{s}$ is the radial
scale length, and $\rho_0$ is provided by
\begin{equation}
\rho_0=\frac{M_\text{h}}{4\pi {R_\text{h}}^3}
\frac{{C_\text{NFW}}^3}{\ln(1+C_\text{NFW})-C_\text{NFW}/(1+C_\text{NFW})}.
\label{eq:NFW_rho0}
\end{equation}
In Equation~(\ref{eq:NFW_rho0}), $R_\text{h}$ is the cut-off
radius of the halo, $M_\text{h}$ is the halo mass within
$R_\text{h}$, and $C_\text{NFW}$ is the concentration parameter
defined by
\begin{equation}
C_\text{NFW}=R_\text{h}/r_\text{s}.
\end{equation}
As a basic model, we choose $M_\text{h}=5\,M_\text{d}$,
$R_\text{h}=30\,h$, and $C_\text{NFW}=5$ for the halo model.
These choices lead to $r_\text{s}=6$.  Concerning a specific
performance test, the halo mass is changed with the other
quantities being left intact.

We construct the equilibrium disk-halo model described above 
using a software tool called many-component galaxy initializer
(MAGI) \citep{magi18}.  Retrograde stars are introduced with
the same way as that adopted by \citet{zh78} and the parameter
$\eta$, which specifies the fraction of retrograde stars, is
set to be 0.25.  We choose the Toomre's $Q$ parameter \citep{toomre64}
to be 1.2 at $R= h$.  In our simulations, the gravitational
constant, $G$, and the units of mass and scale length are
taken such that $G=1$, $M_\text{d}=1$, and $h=1$.

We find from Equation~(\ref{eq:NFWhalo}) that the NFW halo shows
a cuspy density distribution like $r^{-1}$ down to the center.  In
accordance with this characteristic, we adopt Hernquist--Ostriker's
basis set \citep{HO92}.  Because the lowest order members of this
basis set are based on the Hernquist model \citep{lars90} whose
density behaves like an $r^{-1}$ cusp at small radii, that basis
set is suitable to represent the NFW halo with a small number of
expansion terms.  The exact functional forms of the basis set are
shown in the Appendix \ref{appendix:A}.

\subsection{Convergence Tests}\label{subsec:convergence}
\begin{figure}[tbh]
\includegraphics[width=\linewidth]{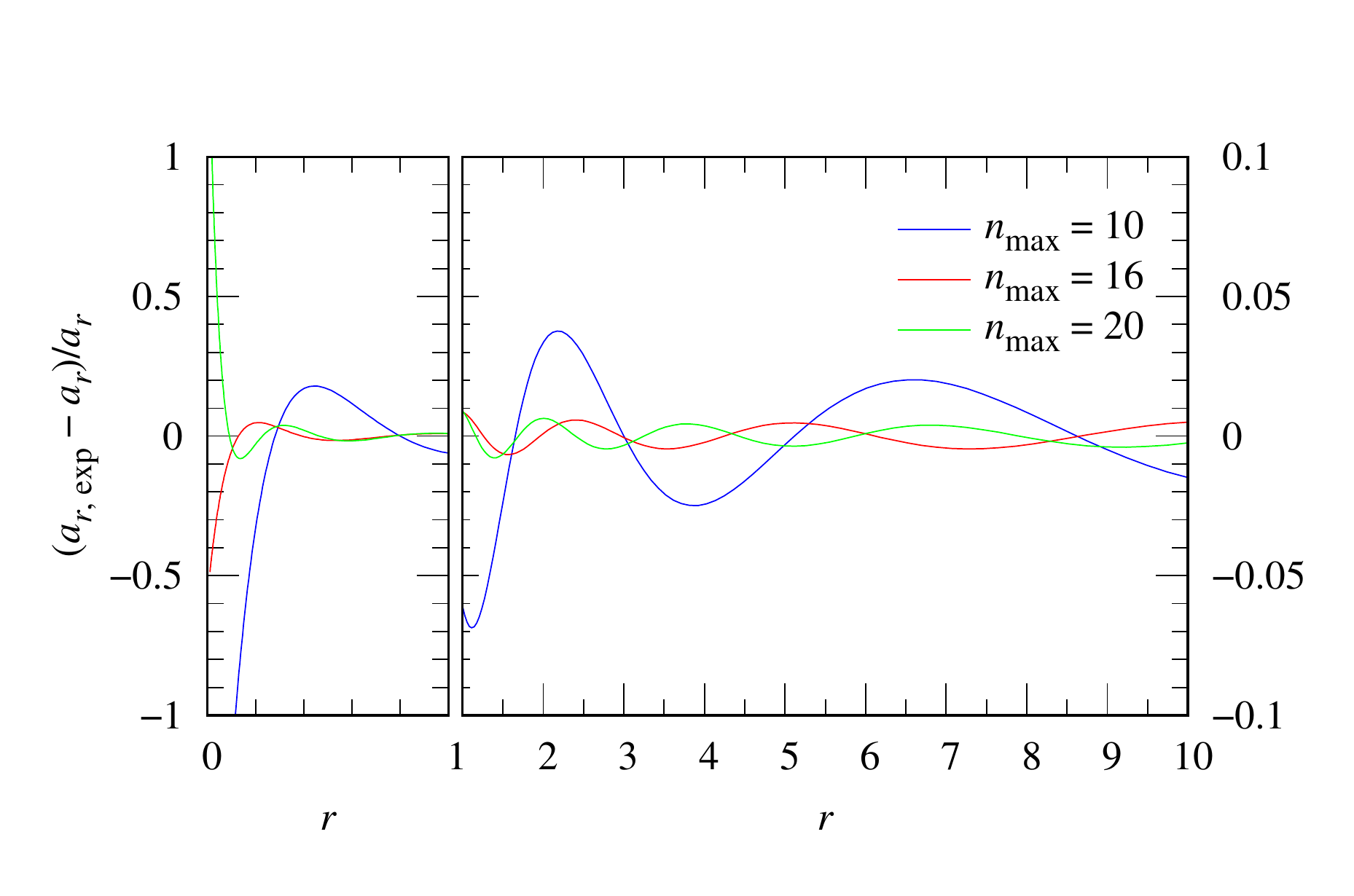}
\caption{Relative radial-acceleration error of the spherically
symmetric NFW halo model as a function of radius.  In this plot,
$a_r$ is the exact acceleration of the NFW halo, while
$a_{r,\text{exp}}$ is the radial acceleration derived from the
expanded potential using Hernquist--Ostriker's basis functions
with the scale length of $a=6$.  The three curves show the
effect of the maximum number of the radial expansion terms,
$n_\text{max}$, on the resulting radial acceleration with the
maximum number of the angular expansion terms, $l_\text{max}=0$,
being retained.  Note that the scaling of the abscissa is changed
at $r=1$ from the left to the right panel, whereby the ordinate
is also re-scaled accordingly.}
\label{fig:expansion}
\end{figure}

\begin{figure*}[htbp]
\centering\includegraphics[width=0.9\linewidth]{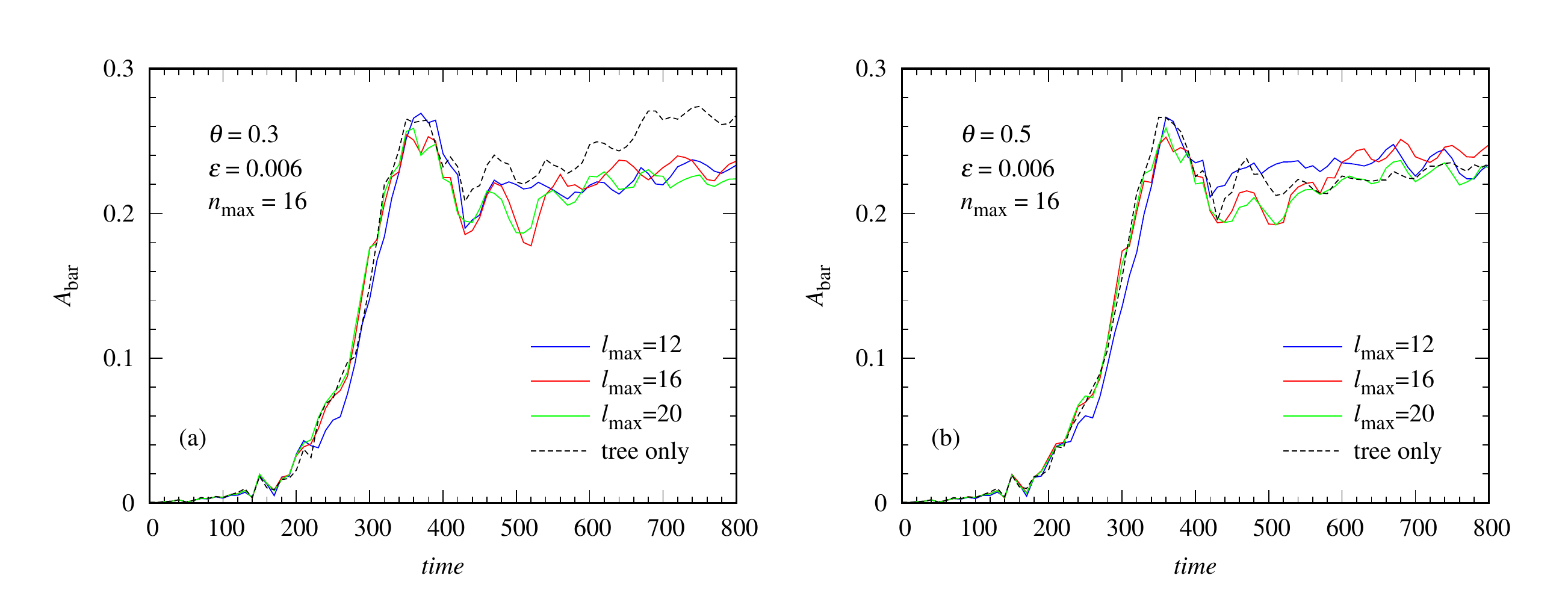}
\caption{Time evolution of the bar amplitude for $l_\text{max}=12, 16$,
and $20$ obtained using the SCF-FDPS code with the opening angle of
$\theta=0.3$ (a), and with that of $\theta=0.5$ (b).  For each value
of $\theta$, the softening length is $\varepsilon=0.006$, and the
maximum number of the radial expansion terms is $n_\text{max}=16$.
As a reference, the corresponding tree code simulation with
$\varepsilon=0.006$ is also plotted for each value of $\theta$.}
\label{fig:baramp}
\end{figure*}

\begin{figure*}[htbp]
\centering\includegraphics[width=0.87\linewidth]{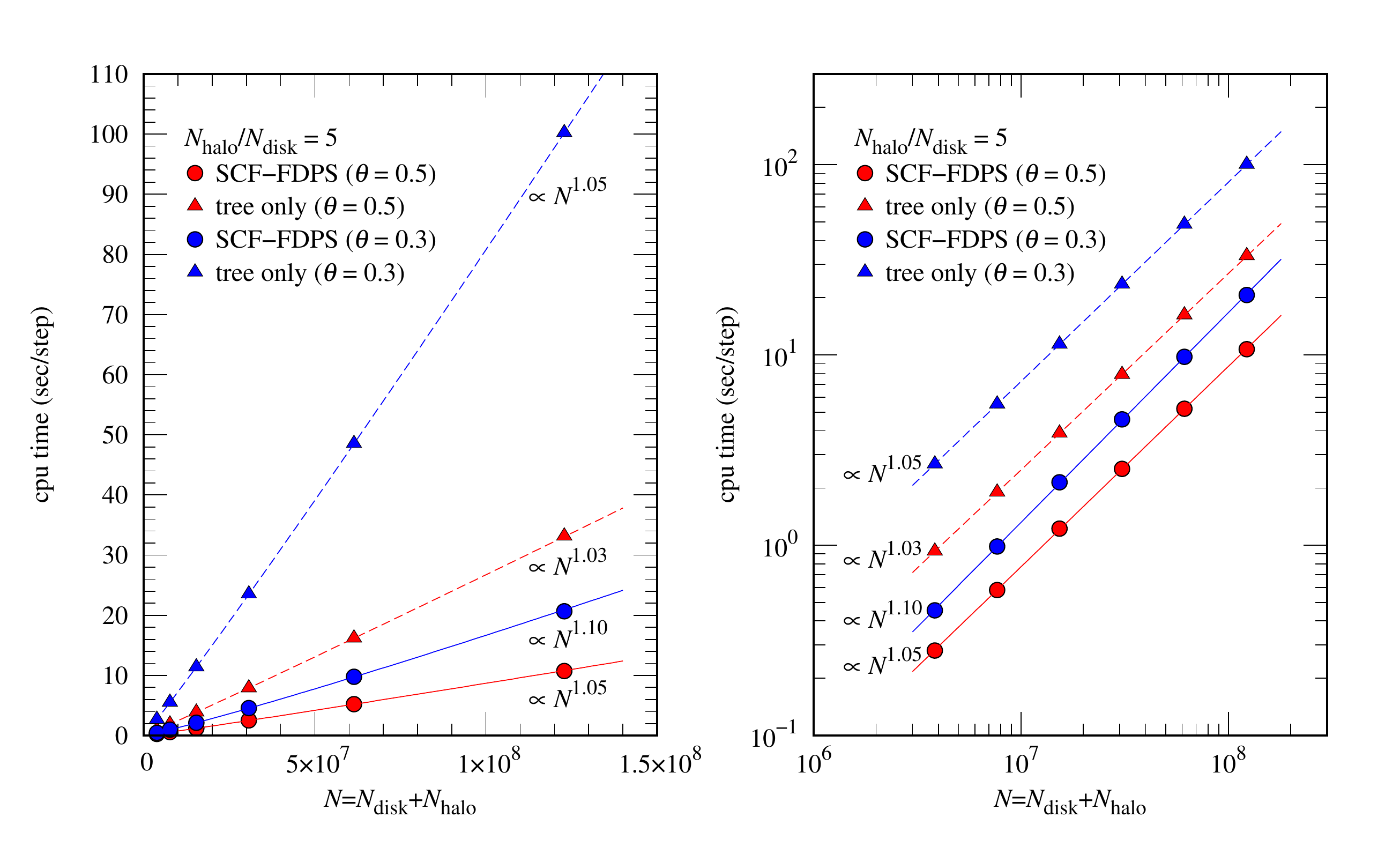}
\caption{Measured cpu time using 64 cores per step in seconds as a
function of the total number of particles, $N=N_\text{disk}+N_\text{halo}$,
where $N_\text{disk}$ and $N_\text{halo}$ are, respectively,
the number of disk particles and that of halo particles with
the ratio of $N_\text{halo}/N_\text{disk}=5$.  The left panel
shows the cpu time on a linear scale, while the right panel
stands for it on a logarithmic scale.  The red symbols represent
the results for $\theta=0.5$, while the blue symbols denote
those for $\theta=0.3$.  The circles display the results
obtained using the SCF-FDPS code, while the triangles exhibit
those using a tree code on which the FDPS library is implemented.
The solid and dashed lines with red and blue colors provide
power-law fits for corresponding data points.}
\label{fig:cpu_N}
\end{figure*}

\begin{figure}[htb]
\centering\includegraphics[width=\linewidth]{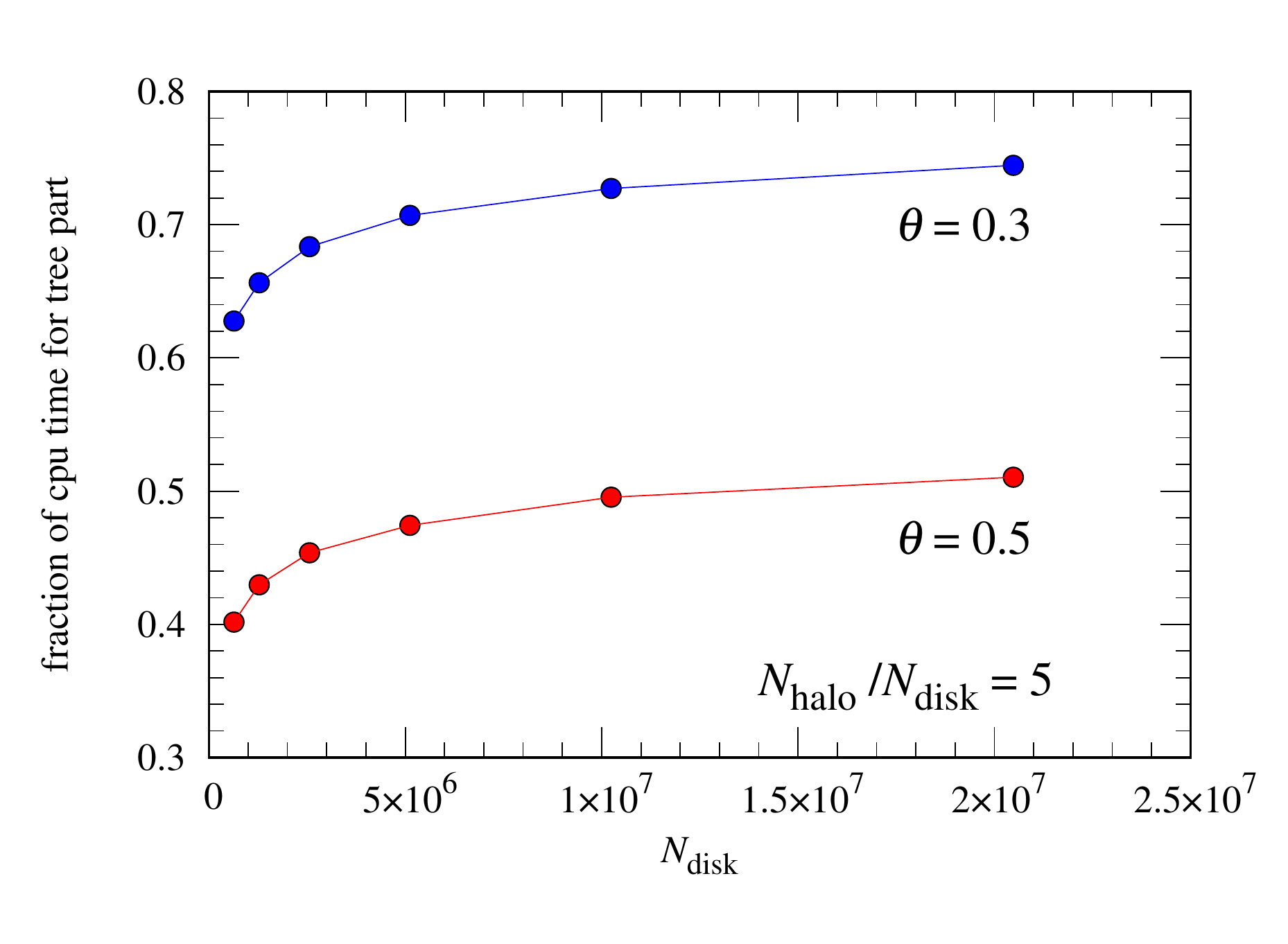}
\caption{Fraction of the cpu time occupied by the tree part
in the SCF-FDPS code as a function of $N_\text{disk}$, which
is calculated from the simulations shown in Figure~\ref{fig:cpu_N}.}
\label{fig:cpu_ratio}
\end{figure}

\begin{figure*}[tbhp]
\centering{\includegraphics[width=\linewidth]{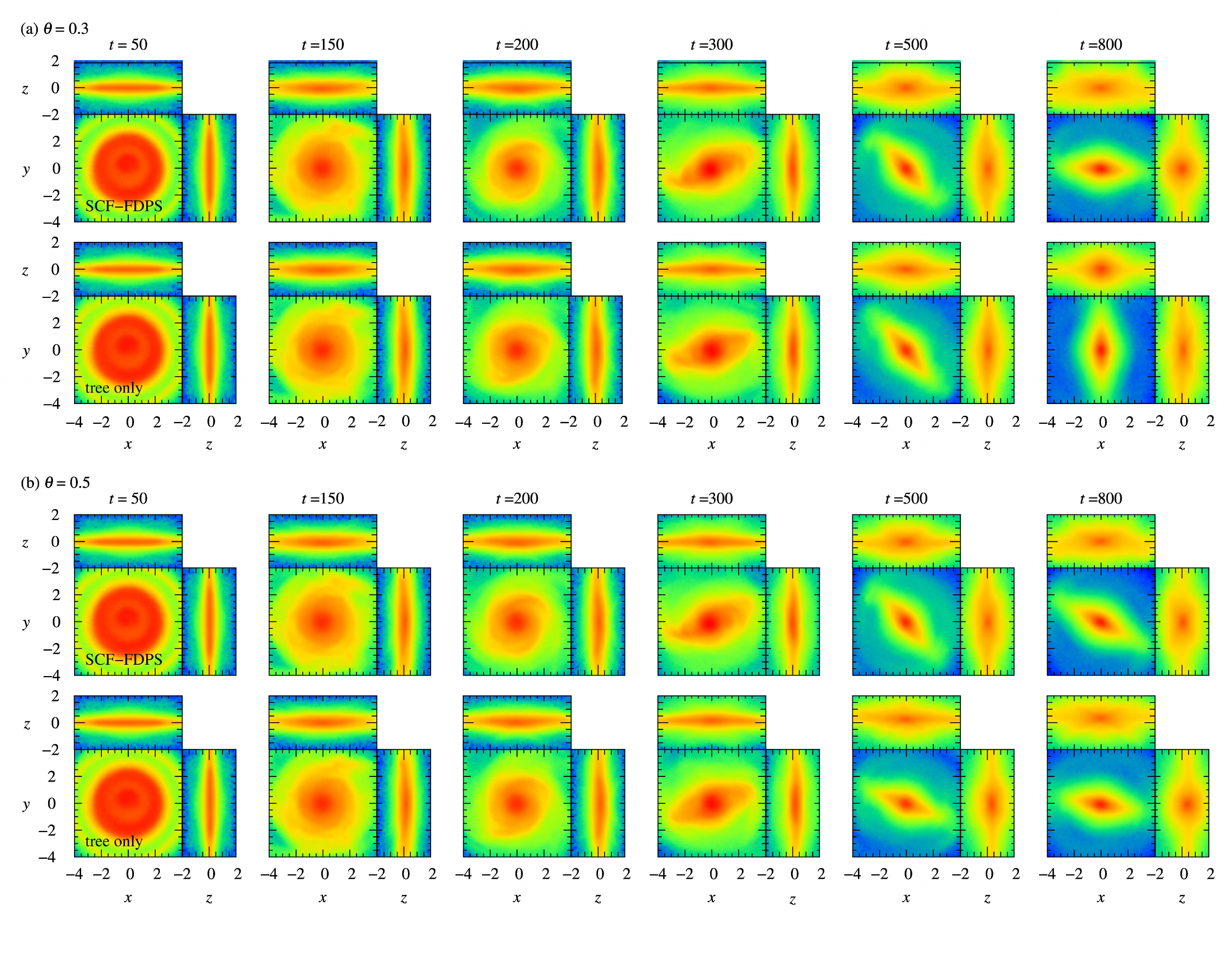}}
\caption{Time evolution of the surface densities of the disk
projected on to the $xy$-, $yz$-, and $zx$-planes for the opening
angle of $\theta=0.3$ (a), and that of $\theta=0.5$ (b).  For each
value of $\theta$, the top panels show the results with the SCF-FDPS
code, while the bottom panels exhibit those with the tree code into
which the FDPS library is implemented.  The softening length is set
to be $\varepsilon=0.006$ for all simulations.  Regarding the SCF-FDPS
simulations, $n_\text{max}=16$ and $l_\text{max}=16$ are used.  Note
that the drift motion along the vertically upward direction is seen
from $t=500$ to $t=800$ for the $\theta=0.5$ simulation with the tree
code.}
\label{fig:contours}
\end{figure*}

For the SCF part in the SCF-FDPS code, we need to specify
$n_\text{max}$ and $l_\text{max}$.  We determine $n_\text{max}$
by comparing the radial acceleration calculated analytically with
that derived from the expanded potential of the spherically symmetric
NFW halo shown in Equation~(\ref{eq:NFWhalo}), which is realized by
retaining $l_\text{max}=0$.  In Figure~\ref{fig:expansion}, the radial
acceleration obtained from the expanded potential for $n_\text{max}=10$,
$16$, and $20$ is compared with the exact one.  The scale length of
the basis functions, $a$, is set to be $a=6$.  This figure indicates
that the radial acceleration obtained with $n_\text{max}=10$ shows
some relatively large deviation from the exact one, while the radial
acceleration with $n_\text{max}=16$ is almost comparable to that with
$n_\text{max}=20$.  From this consideration, we adopt $n_\text{max}=16$.
On the other hand, there is no way to estimate $l_\text{max}$
for a spherical halo model.  To search for an appropriate value
of $l_\text{max}$, we carry out convergence tests in which
$l_\text{max}=12$, $16$, and $20$ are examined with $n_\text{max}=16$
being retained.  We found that the disk-halo model constructed in
Subsection~\ref{subsec:model} forms a bar via the bar instability
(see Figure~\ref{fig:contours}).  Then, we use the time evolution
of the bar amplitude as a measure to determine $l_\text{max}$.

Regarding the parameters related to the tree part, we use $\theta=0.3$
and $0.5$ as an opening angle, and $\varepsilon=0.006$ as a softening
length of the Plummer type.  Gravitational forces are expanded up to
quadrupole order.

We assign $N_\text{disk}=6{,}400{,}000$ to the disk, and
$N_\text{halo}=32{,}000{,}000$ to the halo.  A time-centered
leapfrog algorithm \citep{press86} is employed with a fixed
time step of $\Delta t=0.1$.

For comparison, the same disk-halo model is simulated with a tree code
on which the FDPS library is implemented.  Hereafter, we call this code
the FDPS tree code, which is also applied to the SIMD instructions as
has been done to the SCF-FDPS code.  All the tree parameters are the
same as those employed for the convergence tests.

In Figure~\ref{fig:baramp}, we show the time evolution of
the bar amplitude for $\theta=0.3$ and $0.5$ in each of which
$l_\text{max}=12, 16$, and $20$ are employed, while
$n_\text{max}=16$ is held fixed.  Furthermore, the results
with the FDPS tree code are also plotted.  On the basis of these
results, in particular, paying attention to the behavior of the
exponentially growing phase of the bar amplitude from $t=0$ to
$t\sim 300$, we select $l_\text{max}=16$.

\subsection{Performance Tests}\label{subsec:performance}
We carry out performance tests to examine how fast the SCF-FDPS
code is as compared to the FDPS tree code.  We measure the cpu
time in the cases of $\theta=0.3$ and $0.5$.  For each value of
$\theta$, the Plummer type softening is used with $\varepsilon=0.006$,
and forces are expanded up to quadrupole order.  Again, we use a
time-centered leap-frog method \citep{press86} with a fixed time
step of $\Delta t=0.1$.

In Figure~\ref{fig:cpu_N}, the cpu time using 64 cores per
step is plotted as a function of the total number of particles,
$N=N_\text{disk}+N_\text{halo}$, with the ratio of
$N_\text{halo}/N_\text{disk}=5$ being fixed.  We can see that
the cpu time is nearly proportional to $N$ for both codes, but
that the SCF-FDPS code is at least three times faster than the
FDPS tree code for $\theta=0.5$, while the former is about five to
six times faster than the latter for $\theta=0.3$.  As $N_\text{disk}$
increases, the ratio of the cpu time measured with the FDPS tree code
to that with the SCF-FDPS code decreases for both values of $\theta$.
For example, the ratio is 3.3 for $N_\text{disk}=640{,}000$, while
it is 3.1 for $N_\text{disk}=20{,}480{,}000$ when $\theta=0.5$ is
used.  If $\theta=0.3$ is used, the ratio decreases from 5.9 for
$N_\text{disk}=640{,}000$ to 4.8 for $N_\text{disk}=20{,}480{,}000$.
As Figure~\ref{fig:cpu_ratio} demonstrates, the fraction of the cpu
time exhausted by the tree part in the SCF-FDPS code increases as
$N_\text{disk}$ increases, while the cpu time consumed by the SCF
part is basically proportional to $N_\text{halo}$.  As a result,
that ratio of the cpu time decreases with increasing $N_\text{disk}.$

\begin{figure*}[htbp]
\centering\includegraphics[width=0.87\linewidth]{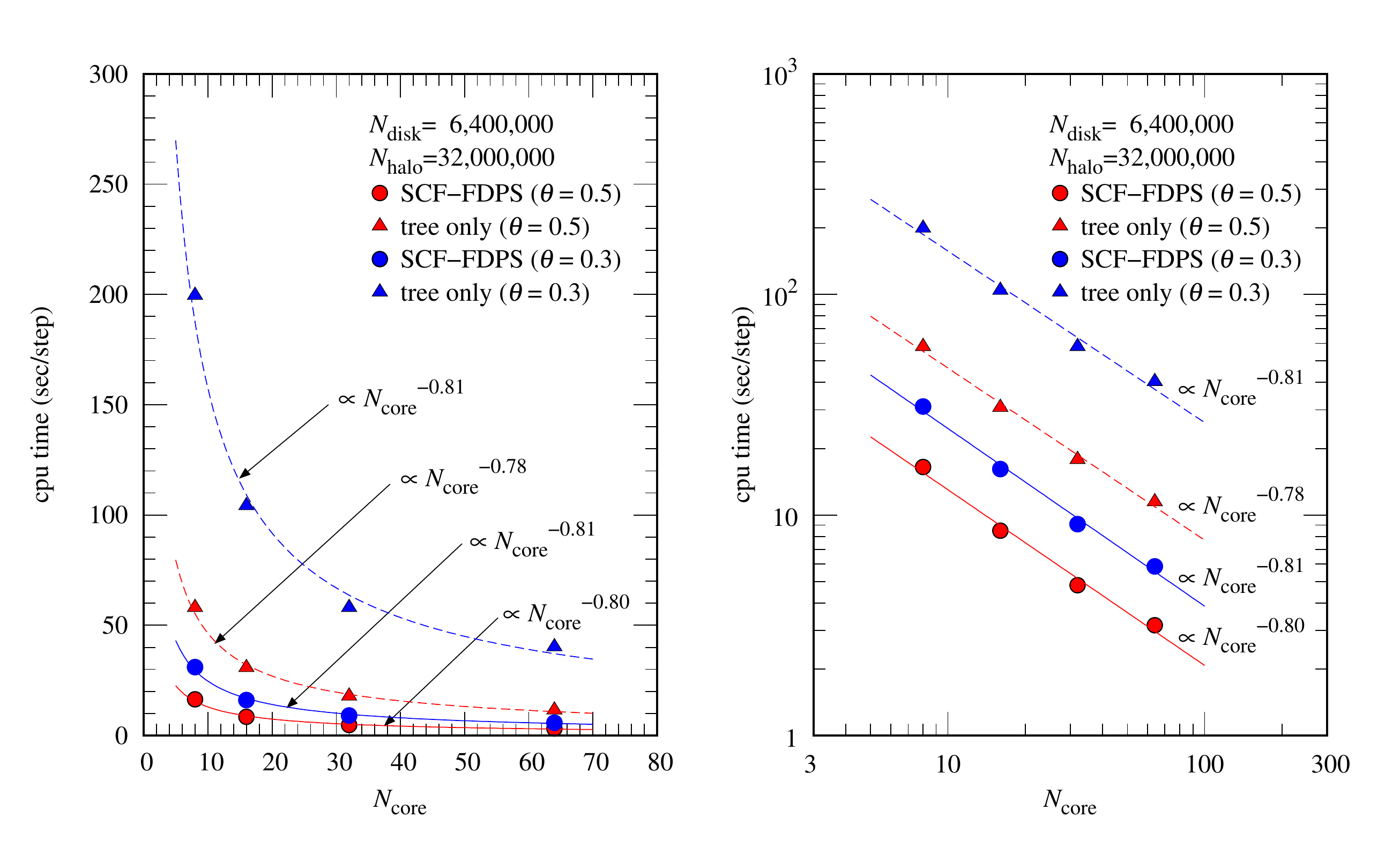}
\caption{Measured cpu time per step in seconds as a function of the
number of cores, $N_\text{core}$.  The number of disk particles is
$N_\text{disk}=6{,}400{,}000$, while that of halo particles is
$N_\text{halo}=32{,}000{,}000$.  The meanings of the symbols and
those of the solid and dashed lines with red and blue colors
are the same as those in Figure~\ref{fig:cpu_N}.}
\label{fig:cpu_Ncore}
\end{figure*}

Next, in Figure~\ref{fig:cpu_Ncore}, the cpu time per step is
plotted as a function of the number of cores, $N_\text{core}$,
used on the computer with $N_\text{disk}=6{,}400{,}000$ and
$N_\text{halo}=32{,}000{,}000$ being unchanged.  Irrespective
of the value of $\theta$, the cpu time scales as
$\sim{N_\text{core}}^{-0.8}$, which means that the cpu time is
almost inversely proportional to $N_\text{core}$ for both codes.
However, the SCF-FDPS code is about 3.6 times faster than the
FDPS tree code for $\theta=0.5$, while the former is approximately
6.4 times faster than the latter for $\theta=0.3$.  In the right
panel of Figure~\ref{fig:cpu_Ncore}, we can see that as $N_\text{core}$
increases, the decrease rate in the cpu time becomes smaller.
This is because the cpu clock is made lowered as $N_\text{core}$
increases.

\begin{figure*}[htbp]
\centering\includegraphics[width=0.87\linewidth]{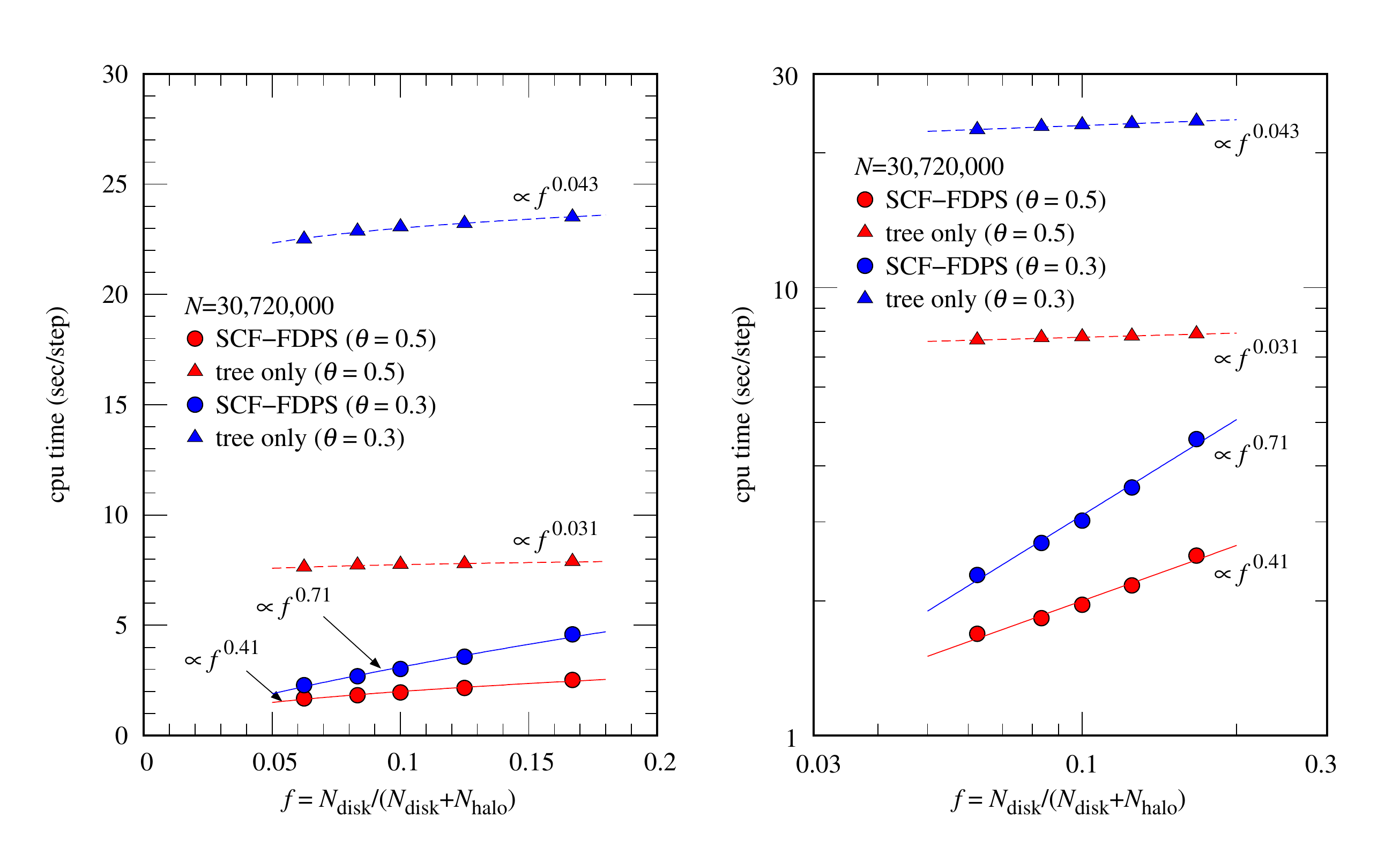}
\caption{Measured cpu time using 64 cores in seconds
per step as a function of the fraction of disk particles,
$f=N_\text{disk}/(N_\text{disk}+N_\text{halo})$.
The total number of particles is $N=30{,}720{,}000$,
and $N_\text{disk}$ and $N_\text{halo}$ are assigned
according to the value of $f$.  In this figure, the
results with $f=1/16, 1/12, 1/10, 1/8,$ and $1/6$ are
plotted.  The meanings of the symbols and those of the
solid and dashed lines with red and blue colors are the
same as those in Figure~\ref{fig:cpu_N}.}
\label{fig:cpu_tree}
\end{figure*}

Last, in Figure~\ref{fig:cpu_tree}, the cpu time using
64 cores per step is plotted as a function of the fraction
of disk particles, $f=N_\text{disk}/N$, where
$N=N_\text{disk}+N_\text{halo}$, and we use $f=1/16$,
$1/12$, $1/10$, $1/8$, and $1/6$.  In this performance
test, we change the ratio of $N_\text{halo}/N_\text{disk}$,
while making the total number of particles unchanged as
$N=30{,}720{,}000$.  As a result, the mass ratio of
$M_\text{halo}/M_\text{disk}$ is not constant but changes
identically to the ratio of $N_\text{halo}/N_\text{disk}$.
The other parameters such as $R_\text{h}$ and $C_\text{NFW}$
are left unchanged.  After all, each halo model specified
by the value of $f$ is constructed by adjusting the value
of $\rho_0$ in Equation (\ref{eq:NFW_rho0}) to the given
$M_\text{halo}$.  Figure~\ref{fig:cpu_tree} indicates how
the fraction of the tree part in the SCF-FDPS code affects
the cpu time.  As a reference, we plot the results using the
FDPS  tree code.  For these tree-code simulations, all particles
are obviously calculated with a tree algorithm, so that the cpu
time may be expected to be independent of $f$.  In reality, the
cpu time depends weakly on $f$, and it is proportional to $f^{0.043}$
for $\theta=0.3$, and to $f^{0.031}$ for $\theta=0.5$.  On the other
hand, the cpu time increases with $f$ if the SCF-FDPS code is used
for both values of $\theta$.  However, for $\theta=0.5$, the SCF-FDPS
code is about 4.5 times faster at $f=1/16$ and about 3.1 times faster
at $f=1/6$ than the FDPS tree code, while for $\theta=0.3$, the former
is about an order of magnitude faster at $f=1/16$ and about 5.1 times
faster at $f=1/6$ than the latter.

\subsection{Simulation Results}
We carry out simulations of the disk-halo system described by
Equations~(\ref{eq:disk}) and (\ref{eq:NFWhalo}) to examine
to what degree the simulation results obtained with the SCF-FDPS
code are similar to those with the FDPS tree code.  The simulation
details are taken over from those adopted for the performance tests.
For each value of $\theta$, the energy was conserved to better than
0.028\% using the SCF-FDPS code, while it was conserved to better
than 0.037\% using the FDPS tree code.  Figure~\ref{fig:contours}
shows the time evolution of the surface densities of the disk projected
on to the $xy$-, $yz$-, and $zx$-planes for $\theta=0.3$ and $0.5$.
We find from this figure that the time evolution of the disk surface
densities obtained with the SCF-FDPS code is in excellent agreement
with that using the FDPS tree code for both values of $\theta$ at
least until $t=500$.  At later times, owing to the difference in
the bar pattern speed from simulation to simulation, the bar phase
differs accordingly.  Even though a difference in the bar pattern
speed is slight at the bar formation epoch, it accumulates with
time, so that the difference in the bar phase becomes larger and
larger as time progresses.  At any rate, the time evolution of the
disk is satisfactorily similar between the two codes.

\begin{figure}[tb]
\centering\includegraphics[width=\linewidth]{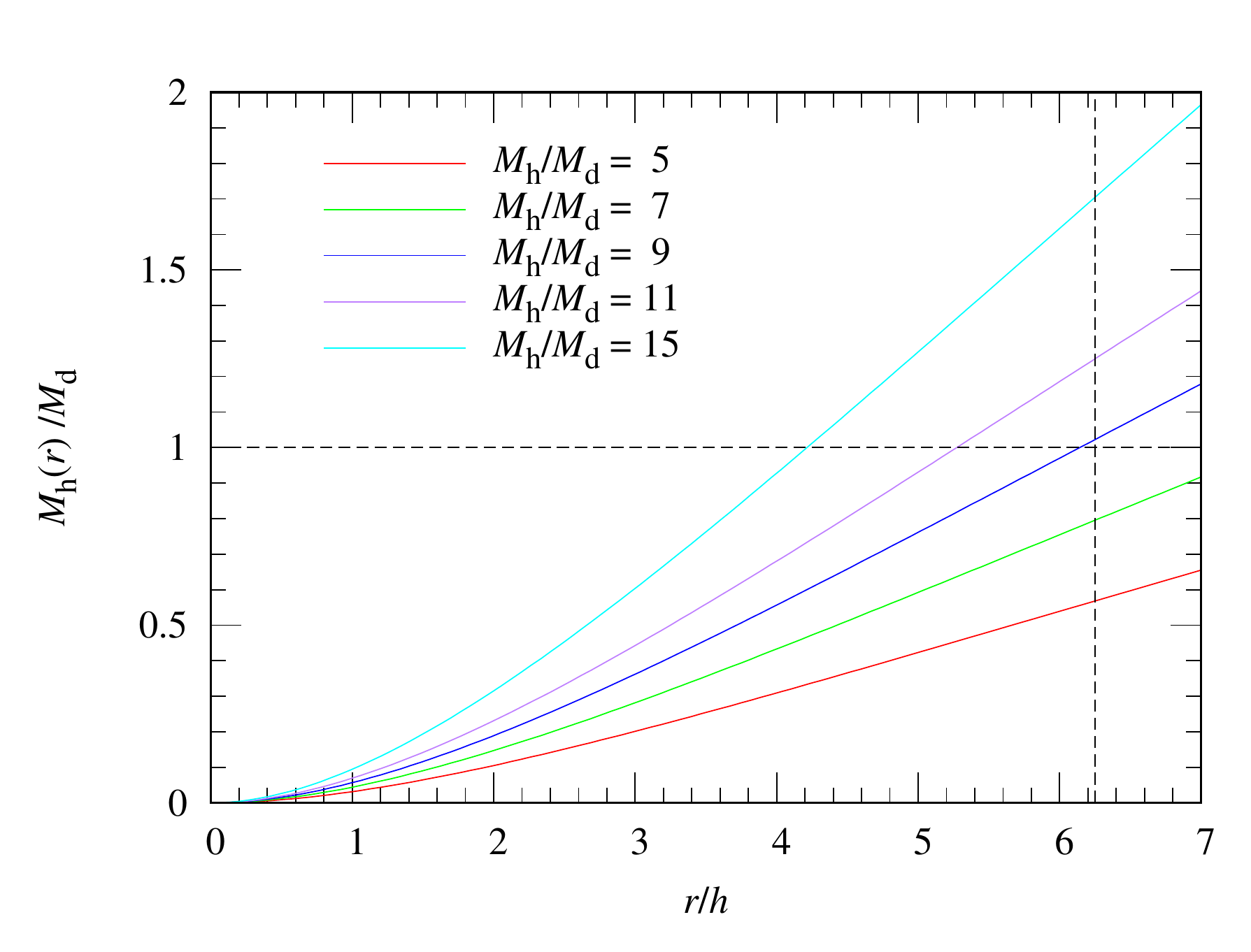}
\caption{Cumulative mass of each halo used in Figure~\ref{fig:cpu_tree}
as a function of radius.  The halo masses are normalized by the disk mass,
and $M_\text{h}=5, 7, 9, 11$, and $15$ correspond to $f=1/6, 1/8, 1/12$,
and $1/16$, respectively, where $f$ is the fraction of disk particles
as in Figure~\ref{fig:cpu_tree}.  The vertical dashed line indicates
the radius of 15 kpc when the radial scale length of the disk is
assumed to be $h=2.4$ kpc.}
\label{fig:halo_mass}
\end{figure}

\section{Discussion}\label{sec:discussion}
We have shown in Figure~\ref{fig:cpu_tree} that the cpu time taken
with the SCF-FDPS code increases as the fraction of $N_\text{disk}$
increases.  In that figure, the mass of each halo is assigned to
that included within $r=30$.  However, if the optical edge of the
disk is about $15$ kpc, this radius corresponds to $r=6.25$ because
the disk scale length is estimated to be $2.4$ kpc \citep{bg16}.
In this case, Figure~\ref{fig:halo_mass} indicates that the halo
mass within $r=6.25$ is at most about 1.7 times the disk mass
even for the largest ratio of $M_\text{h}/M_\text{d}=15$.  Since
the halo mass within the optical edge of the disk is at least
comparable to the disk mass, we may be allowed to regard $f=1/16$
in Figure~\ref{fig:cpu_tree} as a reference value of $f$.  Thus,
if we are based on the results obtained from the simulations with
the value of $f=1/16$ for $\theta=0.3$ and $\theta=0.5$, it follows
that in a practical sense, the SCF-FDPS code is about an order of
magnitude faster than the FDPS tree code for $\theta=0.3$, and that
the former is about 4.5 times faster than the latter for $\theta=0.5$.

We notice from Figure~\ref{fig:contours}(b) that in the simulation
for $\theta=0.5$ executed with the FDPS tree code, the disk begins
to drift upward along the $z$ axis at $t\sim 300$, which continues
to the end of the run, while in the corresponding simulation with
the SCF-FDPS code, no upward drift occurs during the run.  As found
from Figure~\ref{fig:contours}(a), such an upward drift does not
arise in the simulation for $\theta=0.3$ with both codes.  Thus,
in general, tree-code simulations of a disk-halo system do not
necessarily lead to a vertical drift motion of the disk.  Indeed,
in general, linear momentum is not conserved intrinsically in an
exact sense for numerical codes based on expansion techniques such
as tree and SCF codes.  However, our results may suggest that owing
to the small fraction of the tree-based calculation, the SCF-FDPS
code can easily conserve the linear momentum of each component
better than the FDPS tree code to some satisfactory degree.

In our test simulations, we have adopted the softened gravity
due to the softening of the Plummer type because it is easily
implemented in the SCF-FDPS code.  However, in some situations,
spline softening \citep{hk89} may be useful because the force
law turns into the pure Newton's law of universal gravitation
at inter-particle distances larger than twice the softening
length.  Then, we have also implemented the spline softening
in the SCF-FDPS code.

For the SCF part in the SCF-FDPS code, we have used
Hernquist--Ostriker's basis set on the ground that it
well-describes a cuspy density distribution which the
halo model chosen here shows.  In addition, we have also
implemented Clutton-Brock's basis set \citep{cb73}.  This
is suitable for cored density distributions, because the
lowest order members of the basis functions are based on
the Plummer model \citep{Plummer11}.  Therefore, the SCF-FDPS
code can accommodate a wide variety of halo profiles.

In the SCF-FDPS code, disk particles are treated with a tree
algorithm, so that the gas component can easily be included
by implementing an SPH method \citep{gm77,lucy77}, as was
done by \citet{hk89} who named the code TREESPH.  Fortunately,
the FDPS library supports the implementation of an SPH method
by supplying its sample code.  Furthermore, an individual time
step method \citep[e.g.,][]{steve86,hk89,makino91} can also
be set in the SCF-FDPS code, which enables us, for example, to
properly trace particles moving closely around a super-massive
black hole residing at the disk center.  Accordingly, we will
be able to cope with various problems involved in disk galaxies
by equipping additional functions such as SPH and individual
time step methods with the current SCF-FDPS code.

\section{Conclusions}\label{sec:conclusions}
We have developed a fast $N$-body code for simulating disk-halo
systems by incorporating an SCF code into a tree code.  In particular,
the success in achieving the high performance consists in reducing
the time-consuming tree-dependent force calculation only to the
self-gravity of disk particles by applying an SCF method to the
calculation of the gravitational forces between disk and halo
particles as well as that of the self-gravity of halo particles.
In addition, the SCF-FDPS code has the characteristics that the
cpu time is almost proportional to the total number of particles
for the fixed number of cores and almost inversely proportional
to the number of cores equipped on a computer for the fixed number
of particles.  As a result, for a disk-halo system, the SCF-FDPS
code developed here is at minimum about three times faster
and in some case up to an order of magnitude faster, depending
on the opening angle, $\theta$, used in the tree method, and
on the fraction of tree particles,
$f=N_\text{disk}/(N_\text{disk}+N_\text{halo}$), than
a highly tuned tree code like the FDPS tree code.  Of
course, the SCF-FDPS code leads to the time evolution
of a disk-halo system similarly to that with the FDPS
tree code.

We have implemented Clutton-Brock's basis set suitable for cored
density distributions as well as Hernquist--Ostriker's basis set
appropriate for cuspy density distributions on the SCF-FDPS code,
so that it is capable of coping with a wide variety of halo profiles.
Furthermore, because the spline softening as well as the Plummer
softening have been implemented on that code, it will be able to
be applied to the investigation of extensive dynamical problems
of disk-halo systems.

We can easily incorporate both SPH and individual time step methods
into the tree part in the SCF-FDPS code.  Therefore, the SCF-FDPS
code will be able to be extended so that we can tackle central
issues of disk-galaxy simulations like the evolution of a disk
galaxy harboring a central super-massive black hole including
a gas component with a huge number of particles by utilizing
its high performance.

\begin{acknowledgments}
We are grateful to Dr.~Yohei Miki for his advice about the usage
of MAGI.  SH thanks Prof.~Lars Hernquist for his comments on the
manuscript.  This work was supported by JSPS KAKENHI Grant Number
JP21K03626.  Some of the SCF and tree-code simulations were carried
out on the Cray XC50 system at the Center for Computational Astrophysics
at the National Astronomical Observatory of Japan.
\end{acknowledgments}

\begin{appendix}
\section {The density and potential basis functions}\label{appendix:A}
The basis set adopted here is that constructed by \citet{HO92}.
The density and potential basis functions, expressed by
$\rho_{nlm}(\bm r)$ and $\Phi_{nlm}(\bm r)$, respectively,
are represented by
\begin{equation}
\rho_{nlm}(\textbf{\textit{r}})=K_{nl}\frac{M}{2\pi a^3}
\frac{(r/a)^l}{(r/a){(1+r/a)}^{2l+3}}C_{n}^{(2l+3/2)}(\xi)
\sqrt{4\pi}\,Y_{lm}(\theta,\,\phi)
\label{eq:density_basis}
\end{equation}
and
\begin{equation}
\Phi_{nlm}(\textbf{\textit{r}})=-\frac{GM}{a}\frac{(r/a)^l}{(1+r/a)^{2l+1}}
C_{n}^{(2l+3/2)}(\xi)\sqrt{4\pi}\,Y_{lm}(\theta,\,\phi),
\label{eq:potential_basis}
\end{equation}
where $M$ is the mass of the system, $a$ is the scale length,
$C_n^{(\alpha)}(\xi)$ are the ultraspherical, or Gegenbauer
polynomials \citep{as72} with $\xi$ being the radial transformation
defined by
\begin{equation}
\xi=\frac{r-a}{r+a},
\end{equation}
and $Y_{lm}(\theta,\,\phi)$ are spherical harmonics which are
related to associated Legendre polynomials, $P_{lm}(x)$, by
\begin{equation}
Y_{lm}(\theta,\,\phi)=\sqrt{\frac{2l+1}{4\pi}\frac{(l-m)!}{(l+m)!}}
\,P_{lm}(\cos\theta)\exp(im\phi),
\end{equation}
where $i$ is the imaginary unit.

In Equation~(\ref{eq:density_basis}), the normalization factor,
$K_{nl}$, is provided by
\begin{equation}
K_{nl}=\frac{1}{2}n(n+4l+3) + (l+1)(2l+1).
\end{equation}

\end{appendix}
\bibliography{scf_fdps}{}
\bibliographystyle{aasjournal}
\end{CJK*}
\end{document}